\documentclass[times, 10pt,twocolumn]{article}
\usepackage{latex8}
\usepackage{times}
\usepackage{graphics} 
\usepackage{graphicx}
\usepackage{color}
\usepackage{url}      
\usepackage{paralist}

\begin{document}

\title{Technical report: Linking the scientific and clinical data with KI2NA-LHC\thanks{The research and development presented in this paper have 
been supported by the KI2NA project, jointly funded by Fujitsu Laboratories 
Limited, IDA (Irish Industrial Development Agency) and SFI (Science Foundation 
Ireland).}}

\author{V\'it Nov\'a\v{c}ek\\
Digital Enterprise Research Institute\\
National University of Ireland Galway\\
IDA Business Park, Dangan, Galway, Ireland\\
\url{vit.novacek@deri.org}\\
\and
Aisha Naseer\\
Fujitsu Laboratories of Europe Limited\\
Intelligent Society Platform Research Division\\
Hayes Park Central, Hayes, UB48FE, UK\\
\url{Aisha.Naseer@uk.fujitsu.com}\\
}

\maketitle
\thispagestyle{empty}

\begin{abstract}
We introduce a use case and propose a system for data and knowledge
integration in life sciences. In particular, we focus on linking clinical
resources (electronic patient records) with scientific documents and data
(research articles, biomedical ontologies and databases). Our motivation is
two-fold. Firstly, we aim to instantly provide scientific context of
particular patient cases for clinicians in order for them to propose
treatments in a more informed way. Secondly, we want to build a technical
infrastructure for researchers that will allow them to semi-automatically
formulate and evaluate their hypothesis against longitudinal patient data.
This paper describes the proposed system and its typical usage in a
broader context of KI2NA, an ongoing collaboration between the DERI research
institute and Fujitsu Laboratories. We introduce an architecture of the
proposed framework called KI2NA-LHC (for Linked Health Care) and outline the
details of its implementation. We also describe typical usage scenarios and
propose a methodology for evaluation of the whole framework. The main goal of
this paper is to introduce our ongoing work to a broader expert audience. By 
doing so, we aim to establish an early-adopter community for our work and 
elicit feedback we could reflect in the development of the prototype so that 
it is better tailored to the requirements of target users. 
\end{abstract}


\Section{Introduction}\label{sec:intro}



Health care presents a huge segment of the world economy and currently faces tremendous productivity challenges that are in no small part related to the recent data explosion in the related fields.  
The health care stakeholders include pharmaceutical and medical product industries, health care providers, staff and patients, each with different interests and incentives. All of them generates vast pools of data, typically disconnected from each other. The future of data-intensive disciplines is in more efficient data sharing and integration~\cite{naseer2013}. 
Interconnecting the life science data repositories makes them much more actionable and useful in practice, as follows from the generalisation of Metcalfe's law\footnote{The value of a network is quadratic w.r.t. the number of its nodes.} in the broader context of information networks~\cite{infrules1999}. Combination and aggregation of various types of health care-related data provably leads to increases in productivity, reduction of the cost of health care processes and improvement of clinicians' experience when dealing with the data~\cite{dils2012}. All of that is ultimately beneficial for the most important thing in health care - treatment of the patients.

Examples of heterogeneous health care-related data include (but are not limited to): patient data coming from various EHR management and clinical trial systems, genetic testing vendors, longitudinal studies, epidemiological databases and scientific resources (drug, protein and gene databases, biomedical ontologies, etc.). 
Tighter integration of all these types of data sources
facilitates more informed decision making for medical professionals who require various focused and personalised perspectives on the data related to the cases they currently deal with.
More interlinked biomedical resources are also beneficial for scientists in the context of in-silico research with actual patient data, making certain hypothesis instantly testable without a need for tedious literature reviews and expensive experiments.

The concept of Linked Data~\cite{ldbook2011} can facilitate more efficient (re)presentation and processing of biomedical resources by uniform, standardised handling of the large, dynamic and heterogeneous health care-related datasets.
Linked Data has growing enthusiastic support from industry and academia. Its technical bases are the decentralised and general architecture of the World Wide Web and a simple format called RDF~\cite{RDF}, suited for representation and annotation of globally interlinked data. The increasing number of 
data exploitation techniques provided by the Linked Data community naturally offers unprecedented possibilities also for the biomedical data integration. 


The main contribution of this paper is two-fold. Firstly, we present two 
complementary use cases in biomedical data integration that illustrate 
practical problems currently faced by clinicians and biomedical researchers 
(Section~\ref{sec:usage}). Secondly, we describe an architecture, core 
components and ongoing evaluation of KI2NA-LHC, a system we currently develop
to realise the presented use cases (Section~\ref{sec:lhc}). The system 
builds on several of our recent research projects that are summarised in 
Section~\ref{sec:related} (together with other related approaches). We 
conclude the paper 
in Section~\ref{sec:conclusions}.

\Section{Related Work}\label{sec:related}


Approaches related to our work can be classified in several categories. For
a uniform, simple and extensible representation, storage and processing of
data, we use the Linked Data principles and technologies~\cite{ldbook2011}. 
In particular, we extend the approach to distributed storage and dynamic 
querying of linked data based on 
dataspaces~\cite{Franklin:2005:DDN:1107499.1107502}, as elaborated 
in~\cite{DBLP:conf/icde/UmbrichKPPH12} by our colleagues from the KI2NA 
collaboration. 

In order to extract more complex knowledge patterns from the relatively simple
data, we build on the recent advances in the theory of distributional
semantics~\cite{dm2010}, and, in particular, on our work on distributional
data semantics~\cite{Novacek2011iswc,olbook2013novacek}. We combine this
theoretical
groundwork with domain-specific approaches to data integration in life
sciences~\cite{dils2012} and implement the result following the best practices
in computer-based biomedical systems~\cite{bminf2006shortliffe}.

Regarding the user interface
design and modes of typical user interaction, KI2NA-LHC combines principles of
knowledge-based publication search engines (e.g.,
Textpresso~\cite{textpresso2004}, GoPubMed~\cite{gopubmed2008-short} or
CORAAL~\cite{Novacek2010176-short}) and interactive data visualisation
interfaces (see Exhibit~\cite{huynh2007exhibit}, LODPeas~\cite{lodpeas2012} and
SKIMMR~\cite{skimmr2013iui} for the most relevant ones). 

Finally, concerning
the deployment to end-users and software integration, the KI2NA-LHC framework
is being implemented as a set of 
modules for GNU 
Health~\cite{gnuhealth}, a state of the art system for clinical data 
management. The tight software integration with GNU Health is one of the key 
advantages of KI2NA-LHC, as it 
makes the novel automated services available to practitioners within 
an environment they are already used to. To the best of our knowledge, this is 
a feature missing in all related approaches as of now.


\Section{Use Cases}\label{sec:usage}

The high-level goal of KI2NA-LHC is to enable better integration of clinical data (e.g., electronic health records, longitudinal studies and/or clinical databases) with related information present in scientific resources (e.g., research articles, biomedical databases, ontologies, and corresponding Linked Open Data resources). Better integration allows for more efficient ways of exploring the context related to particular information on bed (clinical) and bench (research) side. In the following, we illustrate typical usage scenarios for KI2NA-LHC concerning both of these aspects (in Sections~\ref{sec:usage.benchtobed} 
and~\ref{sec:usage.bedtobench}, respectively). 

To specifically illustrate the use case throughout the section, we are going to use an example user, Alice. As an intern in a hospital who just finished her entry-level medical education, she is involved in daily clinical practice, but also does biomedical research as a part of further postgraduate education. Alice has specialised in viral infections before, however, she is currently dealing with AIDS patients. Since AIDS is related not only to virology, but also to many other fields of biomedicine (such as pharmacology, immunology or genetics), she often needs to consult a lot of resources outside of her primary expertise and thus presents a type of user who benefits most from the KI2NA-LHC technology.

\SubSection{KI2NA-LHC for Clinicians}\label{sec:usage.benchtobed}




The clinical usage scenario is motivated by adverse drug reactions, which can
have serious consequences both for patient safety~\cite{npp2008} and for 
economical 
impact of the associated health care services~\cite{pirmohamed2004}. Apart of 
their general significance, adverse drug reactions also seem to be a 
substantial risk for AIDS patients undergoing antiretroviral 
therapy~\cite{aids2010adr}. If one wants to prevent an outbreak of such an 
adverse event or manage it once it happens, it is necessary to explore 
possibly large 
amount of resources very quickly in order to minimise the impact on the 
patient.

To illustrate the situation in detail, imagine Alice is treating Bob, a 
recently admitted HIV-positive patient who has just experienced acute AIDS 
onset. After admission and initial check that confirmed high potential for
resistance against antiretroviral monotherapy (i.e., using just one drug), Bob 
has been prescribed Zidovudine/Lamivudine/Abacavir, which is a mix of three
antiretroviral drugs aimed to cope with resistant HIV strains due to 
complementary effects of the particular drugs. 

However, Bob quickly develops lipodystrophy (abnormal transformations and 
shifts of fat tissue in his body). As this fact is put into his patient
record, it gets processed by KI2NA-LHC and Alice can immediately see 
lipodystrophy as a likely adverse effect associated with Abacavir according 
to many clinical studies. When she explores that link further, other 
information related to Abacavir appears, including increased risk of heart
attack which is marked as especially relevant to Bob. Another significant fact
in the related information is genetic screening for the presence of the 
HLA-B*57:01 allele. This is due to the fact that the KI2NA-LHC system 
automatically integrated data from the following resources:
\begin{inparaenum}[(i)]
  \item Bob's patient record which indicates hypertension and thus higher 
  susceptibility to the development of coronary diseases;
  \item Biomedical publications which suggest strong relation between the 
  presence of the HLA-B*57:01 allele and hypersensitivity to Abacavir;
  \item An alert of FDA (U.S. Food and Drug Administration agency) that 
  suggests genetic
  screening for the presence of the HLA-B*57:01 allele in AIDS patients to be
  treated with Abacavir. 
\end{inparaenum}

After a quick study of the few original sources returned by KI2NA-LHC as 
directly related to the case, Alice finds out that the risk of severe adverse 
effects in Bob's case is indeed very high. She performs genetic screening of 
Bob and confirms the suspect allele, therefore she recommends a rapid switch 
to another mixture of antiretroviral drugs that does not contain Abacavir or 
similar substances.

This example shows how a routine update of patient's record with a relatively
minor issue led to an instant serendipitous discovery of a potentially much 
more serious adverse effect that could manifest itself in near future. With
KI2NA-LHC, Alice was not only able to identify that risk in an early stage,
but also to automatically retrieve relevant material, study it in detail and 
propose further steps to confirm the risk and ensure it is remedied. Doing the 
same with the current technologies is not impossible, however, the process
usually involves much more time and manual effort, while also being more 
error- and omission-prone. At a global scale, this leads to sub-optimal health 
care with potentially preventable severe consequences.

Other general benefits of the emerging KI2NA-LHC platform for clinicians like Alice are:
\begin{inparaenum}[(1)]
  \item Immediate access to summaries of cutting-edge scientific results related to particular clinical cases (providing broader context for busy clinicians who cannot read hundreds of papers at a time although it may
  improve their decision making capability).
  \item Semi-automated diagnosis of a disease and retrieval of related treatment information using records or social media `life logs' of patients previously exhibiting similar symptoms (decision support).
  \item Automated alert services (for instance, if a stream of data from a patient suddenly exhibits a pattern previously identified as potentially life-threatening in literature).
\end{inparaenum}


\SubSection{KI2NA-LHC for Researchers}\label{sec:usage.bedtobench}



As Alice is not only an aspiring clinical expert, but also a researcher, she is
concerned about the scientific aspects of AIDS as well. One of her research
interests is the activity of APOBEC family of proteins\footnote{Enzymes that 
help generate protein diversity in mRNA editing, see
\url{http://en.wikipedia.org/wiki/APOBEC} for details.} during the HIV
transcription process. She suspects that certain interleukins, such as IL-27,
may be related to the concerned APOBEC activity, but she is not sure how to
prove it due to her limited experience in genetics. Yet when exploring the
literature using KI2NA-LHC interfaces, she quickly discovers that APOBEC3G,
a specific member of the APOBEC protein family, is frequently related to HIV,
and that its gene expansion is semantically related to IL-27. She can
then fetch the articles directly related to these facts and study them
in detail. The whole process takes less than one minute with KI2NA-LHC, which
saved Alice a lot of time she would need to spend otherwise, either tediously
browsing through heaps of largely irrelevant literature or actually
designing and performing the corresponding experiments.

The general benefits of KI2NA-LHC for the user group consisting primarily of
biomedical scientists and pharmaceutical researchers like Alice are mainly the 
following:
\begin{inparaenum}[(1)]
  \item Facility for semi-automated testing of hypotheses using both
  laboratory/experimental and clinical data.
  \item Automated identification of clinical cases related to a research
  phenomenon (e.g., side effects in patients being treated by an already used
  drug that contains compounds present in a currently researched derivative
  drug);
  \item Tools for symbolic analysis of prevalent trends and patterns in large
  amounts of semi-structured or unstructured patient data.
  \item Identification of a genome pattern which is a cause of disease by
  statistical and symbolic analysis of the ``omics'' data incorporated in
  KI2NA-LHC.
\end{inparaenum}

\Section{Implementation of KI2NA-LHC}\label{sec:lhc}

\begin{figure}[ht]
\scalebox{0.45}{\includegraphics{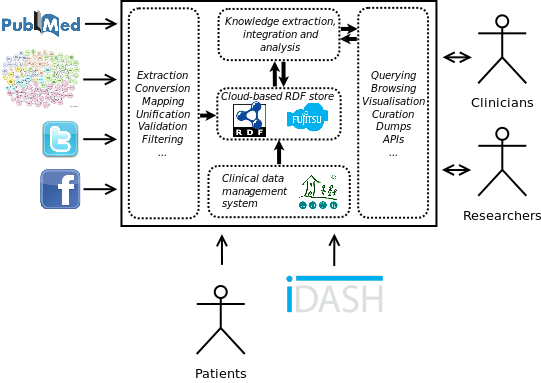}}
\caption{Architecture of KI2NA-LHC}\label{fig:arch}
\end{figure}

The architecture of the KI2NA-LHC framework is depicted in
Figure~\ref{fig:arch}, which gives an overview of the data being processed
by the system, the essential modules and the two general types of expected
users.
Generally speaking, KI2NA-LHC first digests various data related to health care
and biomedical research. It converts them to one uniform format
(RDF~\cite{RDF}), representing everything as binary relationships possibly 
augmented by meta-data like provenance, time, location or certainty. The RDF 
data are stored and indexed in a cloud-based
repository. From the relatively simple statements in the RDF repository, we
compute more expressive knowledge patterns (e.g., semantic similarity,
conceptual clusters of terms, taxonomies or rules). This knowledge is then
served to users via two different user interfaces (one for clinicians, one for
researchers). The users can also provide us with a feedback on the content 
quality which is consequently propagated back into the data store.

In the rest of the section, we provide specific details on the types of data
being ingested by KI2NA-LHC (Section~\ref{sec:lhc.data}), the particular
modules (Sections~\ref{sec:lhc.store}-\ref{sec:lhc.ux}). Last but not least,
we comment on the development and deployment process in
Section~\ref{sec:lhc.devel} and on ongoing evaluation in
Section~\ref{sec:lhc.eval}.

\SubSection{Data and its Pre-Processing}\label{sec:lhc.data}

The data being processed by KI2NA-LHC can be split into several categories:

  {\it 1. Clinical data}: 
  Initially, we are processing sample patient
  data from iDASH, an open repository of real, yet anonymised clinical data
  (cf. \url{http://idash.ucsd.edu/idash-data-collections}; regarding the
  particular data sets, we focus primarily on {\it MT Sample Data}, {\it
  DMITRI Study Data Set} and {\it CDWRnotes}). This allows us to develop and
  test the system with realistic and readily available content without dealing
  with the complex legal and privacy issues usually associated with using raw
  patient data. However, we are able to include arbitrary patient data as they
  become available. This can be easily done through the clinical data
  management system we use as the core platform in KI2NA-LHC (see
  Section~\ref{sec:lhc.devel} for details).

{\it 2. Biomedical research articles}: 
To provide scientific context of
  the processed biomedical data, we employ the Entrez API to PubMed and
  PubMedCentral, open repositories of biomedical abstracts, fulltexts and
  bibliographical information (see \url{http://www.ncbi.nlm.nih.gov/pubmed} and
  \url{http://www.ncbi.nlm.nih.gov/pmc/} for details).

{\it 3. Linked Open Data}: There is an increasing number of freely
  available biomedical resources published as Linked Open Data (cf.
  \url{http://linkeddata.org/}). In particular, we incorporate relevant
  resources offered by the Bio2RDF initiative (cf. \url{http://bio2rdf.org/}),
  and the drug-related data sets listed at
  \url{http://www.w3.org/wiki/HCLSIG/LODD/Data}. In addition to the data
  presently being part of the Linked Open Data cloud, we also process
  content stored in traditional databases (such as the Genome database, cf.
  \url{http://www.ncbi.nlm.nih.gov/genome}) and convert it to RDF/Linked Data
  using the D2R tool developed by our colleagues (cf. \url{http://d2rq.org/}).

{\it 4. Social media}: 
Since a lot of valuable information about patients'
  conditions, subjective assessments and implicitly relevant facts is available
  on popular social networks nowadays, we are going to process that type of
  data as well (focusing primarily on Twitter and Facebook). Initially, we 
  take into account only feeds from pre-defined sets of volunteer user 
  accounts in order 
  asses the relative amount of useful content
  we can get this way. In future (if social media will be deemed a promising
  source of relevant and reliable information), we plan to sieve through
  arbitrary content on the social networks to aggregate population-wide `life
  logs' of clinically relevant information.

As we use the Linked Data principles for uniform storage of the content to be
processed by KI2NA-LHC, we need to convert all data to the RDF format. For the
patient data we use one of the best practices recommended by the
standardisation organisation W3C. In particular, we follow the design pattern
2, use case 3 described in~\cite{RDF-nary}. The pattern allows for 
representing each observation or record about a patient (which has possibly 
multiple facets like
time when taken, value measured, type of the measurement, etc.) as a set of
binary relationships in RDF.

The Linked Open Data resources are already in the RDF format and therefore no
pre-processing is necessary. For natural language text fetched from the
biomedical articles or social networks, we use the co-occurrence analysis and
relation extraction techniques that we describe in~\cite{olbook2013novacek}. 
The result of this process is a CSV file with
$subject,predicate,object,provenance,weight$ records that represent the
extracted relationships together with their provenance (the textual resource
they were extracted from) and confidence weight (how statistically significant
they are). This data is then converted to the observation-based RDF format
mentioned above.

Whenever applicable, the data being processed is mapped to unique
identifiers used in the standard biomedical data sets (as fetched from the
Linked Open Data cloud). This holds especially for entities like drug, gene,
protein or molecule names. The mapping process can utilise the definitions and
synonym extensions of the terms in the linked data sets in order to map their
unique identifiers to lexically similar terms extracted from the less
structured data (e.g., texts or patient records).

\SubSection{Storing and Accessing the Data}\label{sec:lhc.store}

The RDF data produced in the previous step is continuously incorporated
into an in-house RDF store hosted in the Fujitsu Global Cloud (cf.
\url{http://en.wikipedia.org/wiki/Fujitsu_Global_Cloud_Platform}).
The implementation of the cloud-based RDF store is part of the novel data
management infrastructure jointly developed by DERI and Fujitsu.
It provides for scalable and universal data storage and retrieval by
combining the notion of dataspaces~\cite{Franklin:2005:DDN:1107499.1107502}
from the classical databases with the Linked Data principles~\cite{ldbook2011}
and web architecture. The technology builds on the recent research introduced
in~\cite{DBLP:conf/icde/UmbrichKPPH12} by a member of the DERI-Fujitsu team.

\SubSection{Extraction, Integration and Analysis}\label{sec:lhc.k}


The data indexed in the cloud-based RDF store are further analysed by
our framework for emergent knowledge extraction and processing, based on the
research presented in~\cite{Novacek2011iswc,olbook2013novacek}. The framework
makes use of a universal, tensor-based distributional~\cite{dm2010} 
representation of simple
binary statements (essentially a 3-dimensional array of weights associated
with the statements and indices corresponding to the particular arguments of
the statements). This 3D representation can be converted to various 2D matrix
perspectives (i.e., sets of row or column vectors), which can in turn be
analysed by state of the art methods from linear algebra (e.g., vector
comparison or matrix decomposition). Such analysis can discern various
semantic phenomena emerging from the simple data, like:
\begin{inparaenum}[(1)]
  \item implicit similarity relationships between terms;
  \item clusters of similar terms forming concepts;
  \item co-occurrence patterns that can be interpreted as domain-specific
  relationships (such as causality, regulation or expression);
  \item taxonomical hierarchy of the conceptual clusters;
  \item IF-THEN rules.
\end{inparaenum}
The discovered patterns can then be represented as new RDF statements about
the original data and fed back into the central RDF store.

\SubSection{User Interaction}\label{sec:lhc.ux}


Two minimalistic user interfaces are currently being elaborated for KI2NA-LHC
-- one for
clinicians and one for researchers. The clinical interface has to be optimised
for both computers and hand-held devices in order to support the clinicians
everywhere (e.g., even when visiting the patients). It consists of a
simple search box where free-text queries on diseases, symptoms, drugs, genes,
etc. can be entered. The system then fetches all information that relates
to the query from the underlying RDF store, post-processes it via the knowledge
analysis module (ranking the statements according to their relevance), and
displays it to the user in an interactive visualisation. The interface
extends our current tool SKIMMR~\cite{skimmr2013iui}, mostly by providing more
convenient user interaction and additional dynamic visualisations.

The research interface is slightly more complex -- it allows for
graphical formulation of hypothesis in the form of relations between
biomedical entities and their logical combinations. These hypothesis are then
converted into queries and get evaluated against the knowledge stored in
the KI2NA back-end. The result is a numerical assessment of the
plausibility of the hypothesis, as well as an interactive visual summary of the
related information fetched from the back-end (re-using the result presentation
from the clinical interface).

Both interfaces allow users to provide a simple feedback by giving a
`thumbs-up' or `thumbs-down' to particular results. This information then
gets propagated in the back-end knowledge base, improving its quality in time.

\SubSection{Development and Deployment}\label{sec:lhc.devel}


The KI2NA-LHC system is being implemented as a set of extension modules for
the free and open source GNU Health system~\cite{gnuhealth}, which is being
used by many hospitals (especially in developing countries) and also by the
United Nations University. GNU Health serves as a general wrapper for our
back-end and as a basic interface between our data processing components and
the patient records. We also make use of the existing user interfaces in
GNU Health in order to incorporate our user interaction modules into a type of
framework clinicians are used to. This allows us to keep the learning
curve for the new technology feasible, which in turn leads to improved
practical applicability. 

\SubSection{Ongoing Evaluation}\label{sec:lhc.eval}

In order to test the KI2NA-LHC prototype, we are going to recruit sample
users (via the GNU Health community and dissemination at related conferences) 
from the very early stages of development. 
This is to help us in continuous evaluation of the underlying technologies, so
that we can dynamically implement any features implied either by explicit
requests or by the results of the evaluation. 

The ongoing evaluation is two-fold, focusing on quantitative and qualitative
aspects. For quantitative evaluation, we can compare samples of the
automatically computed statements in the KI2NA knowledge base with a golden
standard, for which we primarily use existing biomedical vocabularies
(e.g., MeSH, see \url{http://www.nlm.nih.gov/mesh/} for details). We also need
to build our own gold standard with the sample users, though, in order to
evaluate more complex phenomena and knowledge patterns not captured by 
resources
like MeSH. The comparison with gold standard supports computation of
objective quality measures -- generalised precision and recall, following the
evaluation techniques proposed for tasks like ontology
matching~\cite{euzenat2007b}.

We also address the qualitative evaluation to assess the general applicability
and industrially-relevant performance of the platform. For this
we employ usability surveys based on the standard SUS
methodology (System Usability Scale, cf.
\url{http://en.wikipedia.org/wiki/System_usability_scale}). In addition, we
will produce a clearly defined set of tasks and corresponding results, and
measure performance of our sample users in these tasks (tracking time spent
and results achieved). We will compare their performance when using KI2NA-LHC
and a set of related state of the art solutions (such as 
GoPubMed~\cite{gopubmed2008-short}), which will 
assess the practical contribution of our new system. 


\Section{Conclusions and Future Work}\label{sec:conclusions}



We introduced the KI2NA-LHC framework aimed at data integration and semi-automated knowledge discovery in life sciences. The practical relevance of the framework was illustrated by two realistic use cases involving clinical and research aspects of biomedical data integration. We described the architecture of KI2NA-LHC and outlined the details of its implementation based on our recent research integrated into a state of the art biomedical data management tool, GNU Health. 

KI2NA-LHC is an on-going project that only started in the end of year 2012, 
however, the system builds on a sound basis of previously published work and 
implemented research prototypes. Therefore the bulk of the technical future 
work revolves around software integration of the already available components 
into the architecture scheme presented in this paper. This will make the 
results of our research readily available to end users, which is what matters
most in the area of computer-based medical systems. Apart of the development, 
we need to work on continuous evaluation of the platform's performance. This 
will be done with sample users, following the agile software development 
methodology (i.e., working with users from early stages of the development and 
dynamically incorporating their feedback into the evolving prototype).



\bibliographystyle{latex8}
\bibliography{bib/vit.bib,bib/aisha.bib}

\begin{thebibliography}{10}\setlength{\itemsep}{-1ex}\small

\bibitem{gnuhealth}
{\em GNU Health}.
\newblock Available at \url{http://en.wikibooks.org/wiki/GNU_Health} (Feb
  2013).

\bibitem{npp2008}
{\em National Priorities and Goals: Aligning Our Efforts to Transform
  {America’s} Healthcare}, 2008.
\newblock Available at \url{http://www.pcpcc.net/files/NPP_Report_111809.pdf}
  (Feb 2013).

\bibitem{dm2010}
M.~Baroni and A.~Lenci.
\newblock Distributional memory: A general framework for corpus-based
  semantics.
\newblock {\em Computational Linguistics}, 36(4):673--721, 2010.

\bibitem{dils2012}
O.~Bodenreider and B.~Rance, editors.
\newblock {\em Proceedings of DILS'12 (Data Integration in the Life Sciences)},
  volume 7348 of {\em Lecture Notes in Computer Science}.
\newblock 2012.

\bibitem{gopubmed2008-short}
H.~Dietze et~al.
\newblock {GoPubMed}: Exploring pubmed with ontological background knowledge.
\newblock In {\em Ontologies and Text Mining for Life Sciences}. IBFI, 2008.

\bibitem{euzenat2007b}
J.~Euzenat and P.~Shvaiko.
\newblock {\em Ontology matching}.
\newblock Springer, 2007.

\bibitem{Franklin:2005:DDN:1107499.1107502}
M.~Franklin, A.~Halevy, and D.~Maier.
\newblock From databases to dataspaces: a new abstraction for information
  management.
\newblock {\em SIGMOD Rec.}, 34(4):27--33, Dec. 2005.

\bibitem{ldbook2011}
T.~Heath and C.~Bizer.
\newblock {\em Linked Data: Evolving the Web into a Global Data Space}.
\newblock Synthesis Lectures on the Semantic Web: Theory and Technology. Morgan
  \& Claypool, 2011.

\bibitem{lodpeas2012}
A.~Hogan, E.~Munoz, and J.~Umbrich.
\newblock {LODPeas}: Like peas in a {LOD} (cloud).
\newblock In {\em Proceedings of the Billion Triple Challenge 2012 (at ISWC
  2012)}. Semantic Web Challenge, 2012.

\bibitem{huynh2007exhibit}
D.~F. Huynh, D.~R. Karger, and R.~C. Miller.
\newblock Exhibit: lightweight structured data publishing.
\newblock In {\em Proceedings of {WWW}'07}, pages 737--746, 2007.

\bibitem{RDF}
F.~Manola and E.~Miller.
\newblock {\em RDF Primer}, 2004.
\newblock Available at (November 2008): \url{http://www.w3.org/TR/rdf-primer/}.

\bibitem{textpresso2004}
H.~M. M\"{u}ller, E.~E. Kenny, and P.~W. Sternberg.
\newblock Textpresso: an ontology-based information retrieval and extraction
  system for biological literature.
\newblock {\em PLoS Biology}, 2(11), 2004.

\bibitem{aids2010adr}
M.~Nagpal, V.~Tayal, S.~Kumar, and U.~Gupta.
\newblock Adverse drug reactions to antiretroviral therapy in aids patients at
  a tertiary care hospital in india: A prospective observational study.
\newblock {\em Indian Journal of Medical Science}, 64(6):245--252, 2010.

\bibitem{naseer2013}
A.~Naseer, L.~Laera, and T.~Matsutsuka.
\newblock Enterprise {BigGraph}.
\newblock In {\em Proceedings of the 46th Annual Hawaii International
  Conference on System Sciences (CD-ROM)}. Computer Society Press, 2013.

\bibitem{skimmr2013iui}
V.~Nov\'a\v{c}ek and G.~Burns.
\newblock {SKIMMR}: Machine-aided skim-reading.
\newblock In {\em IUI13 Companion}. ACM, 2013.

\bibitem{Novacek2010176-short}
V.~Nov\'a\v{c}ek, T.~Groza, S.~Handschuh, and S.~Decker.
\newblock {CORAAL}--dive into publications, bathe in the knowledge.
\newblock {\em Journal of Web Semantics}, 8(2-3):176 -- 181, 2010.

\bibitem{olbook2013novacek}
V.~Nov\'a\v{c}ek and S.~Handschuh.
\newblock Empirically grounded emergent knowledge.
\newblock In J.~Voelker and J.~Lehmann, editors, {\em Perspectives of Ontology
  Learning}. IOS Press, 2013.
\newblock In press, pre-print available at \url{http://goo.gl/N2EGa}.

\bibitem{Novacek2011iswc}
V.~Nov\'{a}\v{c}ek, S.~Handschuh, and S.~Decker.
\newblock Getting the meaning right: A complementary distributional layer for
  the web semantics.
\newblock In {\em Proceedings of {ISWC'11}}. Springer, 2011.

\bibitem{RDF-nary}
N.~Noy and A.~Rector.
\newblock {\em Defining N-ary Relations on the Semantic Web}, 2006.
\newblock Available at (June 2008):
  \url{http://www.w3.org/TR/swbp-n-aryRelations/}.

\bibitem{pirmohamed2004}
M.~Pirmohamed, S.~James, S.~Meakin, C.~Green, A.~K. Scott, T.~J. Walley,
  K.~Farrar, B.~K. Park, and A.~M. Breckenridge.
\newblock Adverse drug reactions as cause of admission to hospital: prospective
  analysis of 18,820 patients.
\newblock {\em BMJ}, 329(7456):15--19, 7 2004.

\bibitem{infrules1999}
C.~Shapiro and H.~R. Varian.
\newblock {\em Information Rules: A Strategic Guide to the Network Economy}.
\newblock Harvard Business Press, 1999.

\bibitem{bminf2006shortliffe}
E.~Shortliffe and J.~Cimino, editors.
\newblock {\em Biomedical Informatics: Computer Applications in Health Care and
  Biomedicine (3rd edition)}.
\newblock Springer, 2006.

\bibitem{DBLP:conf/icde/UmbrichKPPH12}
J.~Umbrich, M.~Karnstedt, J.~X. Parreira, A.~Polleres, and M.~Hauswirth.
\newblock Linked data and live querying for enabling support platforms for web
  dataspaces.
\newblock In {\em ICDE Workshops}, pages 23--28, 2012.

\end{thebibliography}

\end{document}